\documentclass[%
superscriptaddress,
amsmath,amssymb,
aps,
prl, 
floatfix,
reprint
]{revtex4-2}

\usepackage{graphicx,xcolor}
\definecolor{darkblue}{RGB}{0,0,150}
\definecolor{nightblue}{RGB}{0,0,100}

\usepackage[
colorlinks,
citecolor=darkblue,
linkcolor=darkblue,
urlcolor=nightblue]{hyperref}

\usepackage[english]{babel}
\usepackage[babel,kerning=true,spacing=true]{microtype}

\usepackage{physics,amscd,dsfont,wasysym,mathrsfs,mathtools, booktabs}

\usepackage[nameinlink]{cleveref}
\Crefname{figure}{Fig{.}}{Figs{.}}
\Crefname{equation}{Eq{.}}{Eqs{.}}

\usepackage{tikz}
\usepackage{braket}
\usepackage{pgfkeys}

\pgfkeys{
    /panel/.is family,
    /panel,
    label/.estore in=\panellabel,
    label=,
    labelxshift/.estore in=\panellabelxshift,
    labelxshift=-4mm,
    labelyshift/.estore in=\panellabelyshift,
    labelyshift=4mm,
    title/.estore in=\paneltitle,
    title=,
    titlexshift/.estore in=\paneltitlexshift,
    titlexshift=0mm,
    titleyshift/.estore in=\paneltitleyshift,
    titleyshift=0mm,
    graphicsopts/.estore in=\panelgraphicsopts,
    graphicsopts={width=\columnwidth},
    box/.estore in=\panelbox,
    box=\columnwidth,
    legend/.estore in=\panellegend,
    legend=,
    legendxshift/.estore in=\panellegendxshift,
    legendxshift=0mm,
    legendyshift/.estore in=\panellegendyshift,
    legendyshift=0mm
}

\newcommand{\panelimage}[2][]{%
    \pgfkeys{
        /panel,
        label=,
        labelxshift=-4mm,
        labelyshift=4mm,
        title=,
        titlexshift=0mm,
        titleyshift=0mm,
        graphicsopts={width=\columnwidth},
        box=\columnwidth,
        legend=,
        legendxshift=0mm,
        legendyshift=0mm,
        #1
    }

    \edef\panelgraphicsoptsExpanded{\panelgraphicsopts}

    \makebox[\panelbox][l]{%
        \begin{tikzpicture}

            \node[inner sep=0] (img)
                {\expandafter\includegraphics
                 \expandafter[\panelgraphicsoptsExpanded]{#2}};

            \if\relax\detokenize\expandafter{\panellabel}\relax
            \else
                \node[
                    anchor=north west,
                    font=\bfseries\scriptsize
                ] at ([xshift=\panellabelxshift,yshift=\panellabelyshift]img.north west)
                    {(\panellabel)};
            \fi

            \if\relax\detokenize\expandafter{\paneltitle}\relax
            \else
                \node[
                    anchor=south,
                    font=\bfseries\tiny,
                    overlay
                ] at ([xshift=\paneltitlexshift, yshift=\paneltitleyshift]img.north)
                    {\paneltitle};
            \fi

            \if\relax\detokenize\expandafter{\panellegend}\relax
            \else
                \node[
                    anchor=north east,
                    text opacity=1,
                    font=\fontsize{1}{3}\selectfont,
                    inner sep=0.5pt,
                    overlay
                ] at ([xshift=\panellegendxshift,
                       yshift=\panellegendyshift]img.north east)
                    {\panellegend};
            \fi

        \end{tikzpicture}
    }%
}%

\begin{document}

\title{
Graphene Islands as a Platform for Band-Structure Engineering
}

\author{Yonah Mann}
\email{yonahmann@mail.tau.ac.il}
\affiliation{School of Physics and Astronomy, Tel Aviv University, Tel Aviv 69978, Israel
}
\author{Tobias Holder}
\affiliation{School of Physics and Astronomy, Tel Aviv University, Tel Aviv 69978, Israel
}
\date{\today}

\begin{abstract}
Multilayer graphenes have emerged as a versatile avenue towards exploring novel electronic phases of matter.
Recently, nanoscale multilayer graphene islands have been realized experimentally, where the island is less than 50 nm in diameter and distinct in layer configuration from the contact portion of the same graphene sheets, the latter of which act as leads for transport probes. 
Here, we theoretically investigate how the proximity between these islands and their chemically identical environment modifies the electronic structure.
Within the tight-binding approximation, we find that the proximity effect is enhanced by the closely matched momenta between both parts, creating an unusually strong band renormalization which persists as the island size is increased, and only limited by the coherence length.
An effective single-unit-cell Hamiltonian to capture these modifications is derived, expressed in terms of interlayer hopping parameters that scale with the relative island area.
We predict a substantially altered low-energy density of states for 3- and 4-layer graphene islands, demonstrating the tunability of the electronic structure by spatial size.
\end{abstract}

\maketitle

\begin{figure}[t]
    \hspace{-2em}
    \makebox[\columnwidth][l]{%
        \panelimage[box=0.67\columnwidth,
                    label=a,
                    labelxshift=0.25em,
                    labelyshift=1.9em,
                    graphicsopts={width=0.67\columnwidth, trim=0.5cm 0 0 0, clip}]{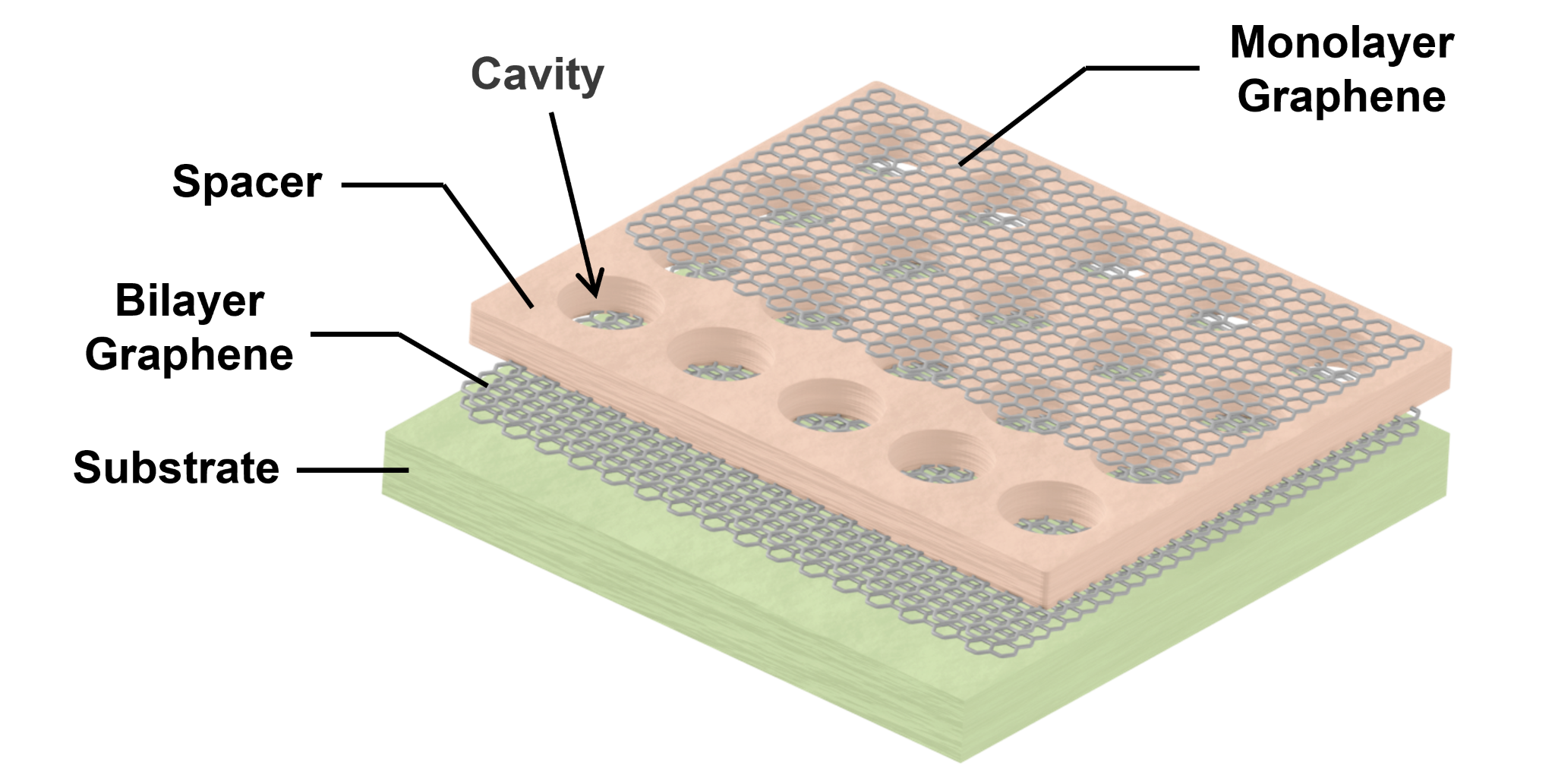}
        \hspace{-2.7em}
        \panelimage[box=0.35\columnwidth,
                    label=c,
                    graphicsopts={width=0.35\columnwidth}]{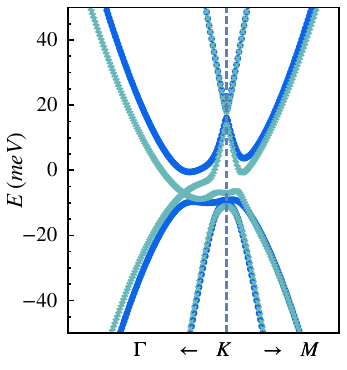}
    }
    \vspace{-1em}
    \panelimage[label=b,labelxshift=0em]{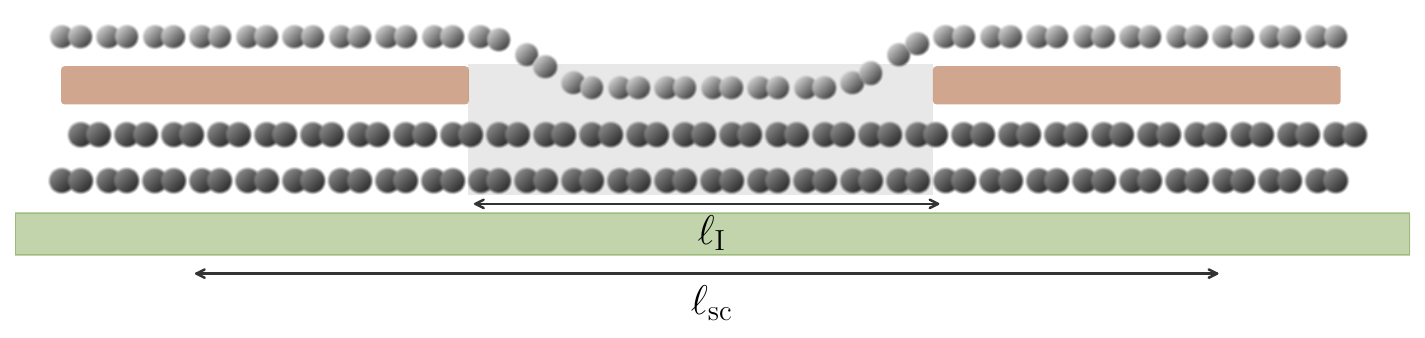}

    \caption{
        \textbf{(a)} Illustration of a device consisting of ABC graphene islands.
                     BLG is placed on top of a substrate.
                     A spacer with cavities etched into it is placed on top of the BLG.
                     MLG is then placed on top, and sags into the cavities forming a trilayer island.
        \textbf{(b)} Side view of a single trilayer island.
                     There are two length scales associated with the islands:
                     $\ell_\text{I}$ - the diameter of the island.
                     $\ell_{\text{sc}}$ - the diameter of the supercell.
        \textbf{(c)} Band-structure comparison between pristine ABA graphene (dark blue) and an ABA graphene island with $\ell_\text{I}/\ell_{\text{sc}} = 12/14$ (light blue).
                     Both calculations set $\Delta_1 = 10 \, meV$.
    }
    \label{fig:figure1}
\end{figure}

\paragraph{Introduction.---}
The fabrication and precision assembly of van der Waals (vdW) materials and heterostructures have in the last few years enabled a plethora of novel non-stoichiometric control parameters which allow one to tune the electronic structure in unprecedented ways~\cite{Andrei2021, Toermae2022a,Nuckolls2024}.
One such control knob is the subtle changes in the electronic structure induced by differences in the atomic coordination in commensurate stacking configurations of graphene~\cite{Zibrov2018,Zhou2021,Zhou2021a,Arp2024,Holleis2025,Zhou2023,Liu2024,Han2025,Auerbach2025,Lu2024}, which is known as \textit{slidetronics} when done in a controlled fashion~\cite{Stern2021,Fei2018,Li2020}.
While the initial macroscopic graphene multilayer devices required preexisting dislocation strips in order for an active switching mechanism to be utilized \cite{Stern2025}, recent work has proposed nanoscale graphene islands as an alternative switchable platform~\cite{Yeo2025,Roy2026}.
In the latter system, two multilayer graphene sheets are separated by an incommensurate superlubricant spacer into which a cavity is etched (cf.~\Cref{fig:figure1}).
Within the cavity, the graphene sheets touch, forming a commensurate single-domain graphene island embedded within the surrounding graphene sheets.
In this letter, we theoretically investigate the resulting effective ground state that is formed inside such nanoscale islands, demonstrating that the spatial size alone serves as an additional tuning parameter. We propose a simple modification of the well-known $k\cdot p$ continuum Hamiltonian ubiquitously used for graphene to efficiently parametrize the influence of the island size.

As our main finding, we report a giant proximity effect in these islands. Normally, proximity effects are constrained to a narrow interfacial boundary due to the exponential decay of the evanescent wave. For the nanoscale islands considered here, we predict a qualitatively different scenario based on the unique environment in which these  islands are embedded. To understand this, we consider the graphene sheets as two parts, the interior graphene region (the island), and a surrounding exterior graphene region that acts as a contact lead (the contacts).
In conventional transport devices with metallic contacts, the large mismatch in crystal momentum between contact and device band structure suppresses hybridization~\cite{Khomyakov2009}.
Similar band-structure modifications of graphene have been reported through coupling to an interfacing material~\cite{Zhou2007,Sundaram2011,Terasawa2023}.
Here, by contrast, both the leads and the system are the same graphene sheets.
Consequently, the hybridization between the island and the contacts is strong and can lead to an appreciable modification of the band structure.
Graphene islands thus provide a distinct device type in which the electronic structure can be tuned continuously via the island size.
Rather uniquely, this tuning knob is potentially accessible without extensive degradation of quasiparticle lifetimes and transport properties.

\paragraph{Length scales.---} \Cref{fig:figure1} depicts a device comprised of multiple graphene islands, where the trilayer graphene (TLG) islands are formed by stacking monolayer graphene (MLG) on top of bilayer graphene (BLG) in rhombohedral coordination (ABC)~\cite{Zhou2021,Zhou2021a,Arp2024,Holleis2025}, which will be our primary analysis example, but similar conclusions hold for ABA~\cite{Zibrov2018,Zhou2023} and ABCA~\cite{Liu2024,Han2025,Auerbach2025} stacking (cf.\ Suppl.\ Info.~\footnote{See Suppl.\ Info.\ for more information, which presents all tight-binding models and parameters, details the derivation and fitting of the effective Hamiltonian, discusses the various effects of modified onsite potentials, as well as additional aspects of the supercell construction~\cite{Hodson2022}.}).
In the following, we assume that the islands are identical and arranged periodically, creating a supercell with closed (periodic) boundary conditions and a well defined band structure.
Such a device is naturally characterized using two length-scales: the island diameter $\ell_I$ and the supercell diameter $\ell_{\text{sc}}$.
The quantum coherence length $\ell_Q$ is the characteristic length scale beyond which spatially separated regions may be treated as effectively separable \cite{Malpetti2016}. We are thus interested in sizes $\ell_{\text{sc}} \lesssim \ell_{Q}$ where different regions inside the supercell can still interfere.
We note that graphene devices typically exhibit a long electronic mean free path~\cite{Mayorov2011,Banszerus2016}, which is believed to be of the same order as $\ell_{Q}$~\cite{Hwang2008,Yankowitz2019}.
Therefore, the hybridization effects at the nanometer scale discussed here will not be limited by $\ell_{Q}$, even for supercells containing hundreds of unit cells.

Using a tight-binding model that accounts for the coupling between the island and contacts, we now show that the electronic wavefunction is not confined to the isla    nd, but instead is distributed between the island and the environment in proportion to their relative areas,
\begin{equation}
    \eta = \frac{A_I}{A_{\text{sc}}} = \left( \frac{\ell_I}{\ell_{\text{sc}}} \right)^2
\end{equation}
This hybridization produces persistent and significant modifications of the band structure that can be parametrized in terms of $\eta$.

\paragraph{Theoretical Model.---}
\label{section:model}

\begin{figure}
    \panelimage[box=0.98\columnwidth,
                label=a,
                labelxshift=-1em,
                labelyshift=0em,
                graphicsopts={width=0.95\columnwidth, trim=1.1em 0.5em 1.1em 0.5em, clip}]{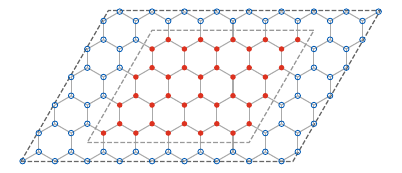}
    \panelimage[label=b,
                labelxshift=-1em,
                labelyshift=0em,
                graphicsopts={width=0.97\columnwidth}]{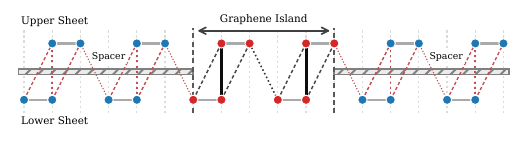}

    \caption{
        \textbf{(a)} In-plane view of the supercell and island regions. Notice that the region boundaries are parallelograms.
                     This is done in order to match the pristine graphene unit-cell geometry.
        \textbf{(b)} Side view depicting the spacer between the sheets in the contacts.
                     Hoppings crossing the spacer are removed, while the interlayer hoppings inside the island are retained.
    }
    \label{fig:tb_model}
\end{figure}

We construct a tight-binding model as illustrated in \Cref{fig:tb_model}.
Denoting the pristine island Hamiltonian as $H_{I_0}$, we define $H_{C_0}$ to be the same as $H_{I_0}$ only with interlayer hopping terms that cross the spacer set to zero.
The supercell Hamiltonian $H_\text{SC}$ can then be expressed as $H_{\text{SC}} = H_I \otimes H_C$, where
$
    H_I = \otimes_{j \in I} H_{I_0},\  H_C = \otimes_{j \in C} H_{C_0}
$
and $I$ ($C$) denote the set of all unit cells within the island and contacts, respectively.
This construction effectively decouples the upper and lower graphene sheets in the contacts.
For $H_\text{SC}$ to be well defined, we must ensure that each primitive unit cell is uniquely assigned to either the island or contact region, with no boundary intersections within the unit cell.
Therefore, we take the supercell and the island shape to be parallelograms, which align with the primitive graphene lattice vectors.

The low energy bands of graphene are located at the $K$ and $K'$ points, and have valley and spin degrees of freedom, which remain degenerate in the single-particle picture~\cite{Neto2009}.
The resulting spinless tight-binding Hamiltonian thus comprises an
$N_{\text{cells}} \times N_\text{orb}$ dimensional Hilbert space, where $N_{\text{cells}}$ is the number of unit cells in the supercell while  $N_{\text{orb}}$ is two times the number of graphene layers.
In our analysis, the hopping amplitudes retained in the contacts are kept equal to those of the pristine island stacking, rather than being replaced by the values appropriate for the contact's stacking.
For example, in a TLG island, the contact consists of MLG and BLG, yet the surviving hopping amplitudes are taken to be those of pristine TLG.
This simplification avoids the introduction of additional effects associated with stacking-dependent hopping terms.
Since the stacking-dependent variations in the hopping amplitudes are small, we do not expect them to qualitatively affect our conclusions~\cite{Note1,Ghazaryan2023}.

\paragraph{Numerical Results.---}
\label{section:numerical_results}

To understand how strongly the island hybridizes with the graphene contacts, we quantify the fraction of the electron wavefunction residing within the island.
The wavefunctions of the tight-binding model are expressed in the basis of the atomic orbitals as 
$
    \ket{\psi_n(k)} = \sum_i c_{n,i}(k) \ket{\phi_i}
$,
where the sum is over all the orbitals in the supercell and $n$ denotes the band index for normalized eigenstates.
The fraction of the wavefunction residing within a specific region is then calculated by summing the weights of the wavefunction for the orbitals belonging to the island as
$
    W_{n,R}(k) = \sum_{i \in R} |c_{n,i}(k)|^2,
$
where $i$ is the orbital index.

\begin{figure}
    \includegraphics{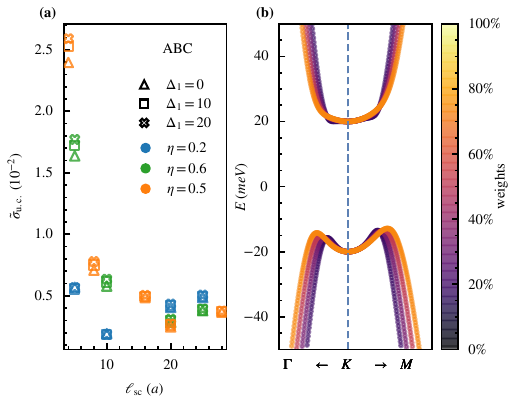}
    \caption{
        \textbf{(a)} Comparison of the normalized standard deviation of the wavefunction weight across the unit cells ($\sigma_{\text{u.c.}}$) of an ABC island for different relative area $\eta$ and applied displacement field $\Delta_1$ (in meV).
        As can be seen, regardless of $\eta$ and $\Delta_1$, the weight of the wavefunction across the different unit cells remains uniform even as the supercell size is increased. $a$ is the lattice spacing.
        \textbf{(b)} Weight of wavefunction sitting inside an ABCA island ($\Delta_1 = 20 \, meV$) for different ratios of $\ell_\text{I} / \ell_{\text{sc}}$: $12/14, 10/14, 8/14, 6/14$.
        The weight is calculated for each momentum independently.
        The weight is completely determined by the relative area of the island to the supercell $\eta$, regardless of stacking configuration and $\Delta_1$ value,
        and does not vary between different momenta near the $K$ point.
    }
    \label{fig:wavefunction_unit_cell_deviation}
    \label{fig:wavefunction_k_space_distribution}
\end{figure}

\Cref{fig:wavefunction_k_space_distribution} depicts the resulting weight of an ABCA island for different momenta near the $K$ point.
Remarkably, the weight is equal to the relative area, and is independent of momentum and band index:
\begin{equation}
    \label{eq:island_weight}
    W_{n,\text{I}}(k) = \eta
\end{equation}
This relation shows that the combined system of island with environment does not exhibit any proximity effect at all, at least not up to the sizes we tested, instead \text{red}{completely hybridizing} \text{green}{it completely hybridizes} the inside and outside wavefunctions, without any signs of evanescent decay of either eigenstate in the converse part of the system.

This finding motivates us to investigate the spatial distribution of the wavefunction more finely.
In order to probe the spatial inhomogeneity, we take the standard deviation of $W_{n,j}(k)$ between unit cells $j$~\cite{Ross2019First}
\begin{align}
    \sigma_n^2(k) &= \frac{1}{N}\sum_{j}\left(W_{n,j}(k) - \overline{W}_n(k)\right)^2,
\end{align}
where $N$ is the number of unit cells in the supercell and $\overline{W}_n(k) = \frac{1}{N}\sum_j W_{n,j}(k)$.
For the sake of conciseness, we further sum the normalized $\sigma_n^2(k)$ for all of the energy bands and momenta \cite{Taylor1997Error},
$
    \tilde\sigma_{\text{u.c.}}^2 = \sum_{n,k} \sigma^2_n(k)/\overline W_n^2(k),
$
where $n$ iterates over all low-energy bands.

\Cref{fig:wavefunction_unit_cell_deviation} depicts $\tilde\sigma_{\text{u.c.}}$ for an ABC island as a function of supercell size $\ell_{\text{sc}}$ for different applied displacement fields and relative island areas $\eta$ at the $K$ point.
Since $\tilde\sigma_{\text{u.c.}}$ stays under $1\%$ once $\ell_{\text{sc}}$ exceeds 5 unit cells, the weight of the wavefunction across the different unit cells comprising the supercell is approximately uniform.
\Cref{eq:island_weight} is an immediate consequence of this result. 
Another test of the uniformity of the wavefunction is the absence of edge states at the boundary of the island.
Repeating the calculation for $W_{n,\text{B}}$, where $B$ denotes the boundary between the island and the contacts of the supercell, shows that it is negligible at all supercell sizes, confirming the absence of any edge modes becoming trapped there.

The strong hybridization effect is not special to ABC, and is indeed found for all stacking configurations (ABA, ABC, and ABCA) and applied displacement fields $\Delta_1$ that we tested~\cite{Note1}.
As a likely explanation of our observations, we note that the decay length of a proximitized wavefunction is given in the WKB approximation as $\lambda=\hbar/\sqrt{2m\Delta E}$, where $m$ is the effective mass and $\Delta E$ is the energy mismatch. A conservative estimate using the bare electron mass and energy shifts by $5$ meV then yields $\lambda\sim 10 a$. Our supercell calculations show that $\lambda$ is at least an order of magnitude larger than that as we were unable to detect any splitting of the eigenvalues into subsystem-specific dispersions near the $K$-point, even for the largest systems we tested ($\ell_{sc}=32$).

\paragraph{Effective Hamiltonian.---}
\label{section:effective_hamiltonian}

In order to understand the behavior of the islands as well as obtain a computationally useful description, we propose an effective tight-binding Hamiltonian for a single unit cell.
Starting from the pristine tight-binding Hamiltonian for the bulk, we proceed by renormalizing the interlayer hopping parameters based on the (spatially dependent) modification of the interlayer hopping amplitudes between island and contact.
The effective Hamiltonian for ABC islands in the ($A_1,B_1,A_2,B_2,A_3,B_3$) orbital basis reads,
\allowdisplaybreaks
\begin{subequations}\label{eq:abc_island}
\begin{align}
    2\Delta_1 &= U_1 - U_3,  \qquad 3\Delta_2 = \frac{U_1 + U_3}{2} - U_2 \label{eq:abc_island:e} \\[0.7em]
    H_{\text{tri},\Delta_1} &= \operatorname{diag}(\Delta_1, \Delta_1, 0, 0, -\Delta_1, -\Delta_1) \label{eq:abc_island:a} \\[0.7em]
    H_{\text{tri},\Delta_2} &= \operatorname{diag}(\Delta_2, \Delta_2,
                              -2\Delta_2, -2\Delta_2,
                              \Delta_2, \Delta_2) \label{eq:abc_island:b} \\[0.7em]
    H_{\text{ABC},0}^{\text{island}} &= 
    \begin{pmatrix}
    0 & v_0 \pi^\dagger & v_4 \pi^\dagger & v_3 \pi & 0 & \frac{1}{2}\gamma_2' \\
    v_0 \pi & \delta & \gamma_1 & v_4 \pi^\dagger & 0 & 0 \\
    v_4 \pi & \gamma_1 & \delta & v_0 \pi^\dagger & v_4' \pi^\dagger & v_3' \pi \\
    v_3 \pi^\dagger & v_4 \pi & v_0 \pi & \delta & \gamma_1' & v_4' \pi^\dagger \\
    0 & 0 & v_4' \pi & \gamma_1' & \delta & v_0 \pi^\dagger \\
    \frac{1}{2}\gamma_2' & 0 & v_3' \pi^\dagger & v_4' \pi & v_0 \pi & 0
    \end{pmatrix} \label{eq:abc_island:c} \\[0.7em]
    H_{\text{ABC}}^{\text{island}} &= H_{\text{ABC},0}^{\text{island}} + H_{\text{tri},\Delta_1} + H_{\text{tri},\Delta_2} \label{eq:abc_island:d},
\end{align}
\end{subequations}
where $U_1$, $U_2$, and $U_3$ are the potentials of the different layers, $\pi = \xi k_x + ik_y$ ($\xi = \pm 1$) is the valley index, $v_i = \frac{\sqrt 3}{2} a \gamma_i$, and $a = 0.246~\mathrm{nm}$ is the lattice constant \cite{Ghazaryan2023}.
As the spacer is situated between layers $2$ and $3$, the modified hopping parameters are $\gamma_1'$, $\gamma_2'$, $\gamma_3'$, and $\gamma_4'$.

\begin{figure}
    \includegraphics[width=.95\columnwidth]{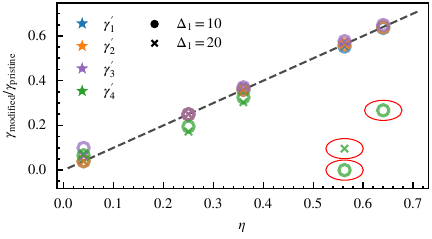}
    \caption{
        Dependence of the fitting parameters for the effective Hamiltonian of an ABC island as a function of $\eta$ for different $\Delta_1$ (in meV).
        The dashed line corresponds to $y=x$.
        $\gamma_1'$, $\gamma_2'$ and $\gamma_3'$ closely follow this relation, independent of $\Delta_1$.
        The deviations observed for $\gamma_4'$ are attributed to the insensitivity of the fitting procedure to changes in $\gamma_4'$. 
    }
    \label{fig:modified_parameters}
\end{figure}

We first extract $\gamma_1'$ and $\gamma_2'$ from the $K$-point energies and then determine $\gamma_3'$ and $\gamma_4'$ by minimizing the deviation between the effective and full-supercell band structures~\cite{Note1}.
\Cref{fig:modified_parameters} depicts the relative value of the modified hopping amplitudes as a function of $\eta$, clearly indicating a linear dependence.
Despite the different microscopic origins of the hopping amplitudes, we arrive at a simple geometric relation between the pristine hopping amplitudes and the modified ones,
\begin{equation}
    \gamma' = \eta \, \gamma.
    \label{eq:modified_vs_pristine_hopping_ratio}
\end{equation}
This relation turns out to be surprisingly general, and holds independently of the applied displacement field as well as for the effective Hamiltonians of ABA and ABCA islands~\cite{Note1}. 
We note that variations in $\gamma_3'$  and $\gamma_4'$ produce only minor changes in the low-energy bands, making these parameters weakly constrained by the fitting procedure, which is the likely cause of the few deviating data points in \Cref{fig:modified_parameters}.
We also tested that the choice of onsite potentials in the island model does not affect any of our conclusions.
Namely, we find that modifying the onsite potentials on the dimer orbitals from the outset in the supercell calculation yields results consistent with fitting them via a free parameter after the fact, without changing the  hopping amplitudes~\cite{Note1}. 

In summary, since the wavefunction is fully hybridized between the graphene island and the graphene contacts, it is natural that  \Cref{eq:modified_vs_pristine_hopping_ratio} should describe the average hopping amplitude between the partially detached layers.

\paragraph{Density of States.---}
\label{appendix:dos}

\begin{figure}
    \vspace{-1.65em}
    \hspace{-2.5em}
        \panelimage[box=1\columnwidth,
                    label=a,
                    labelxshift=0em,
                    graphicsopts={width=1\columnwidth}]{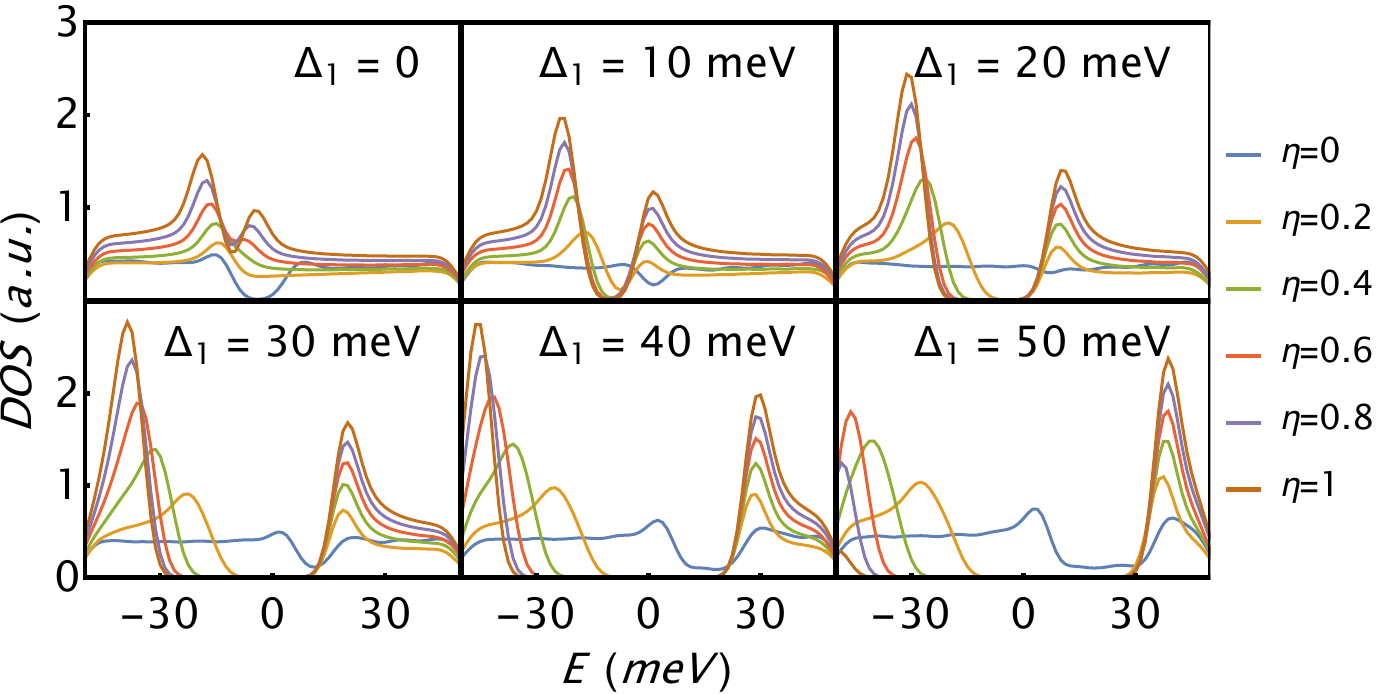}
    
    \vspace{-2em}
    \hspace{-2.5em}
        \panelimage[box=1\columnwidth,
                    label=b,
                    labelxshift=0em,
                    graphicsopts={width=1\columnwidth, clip}]{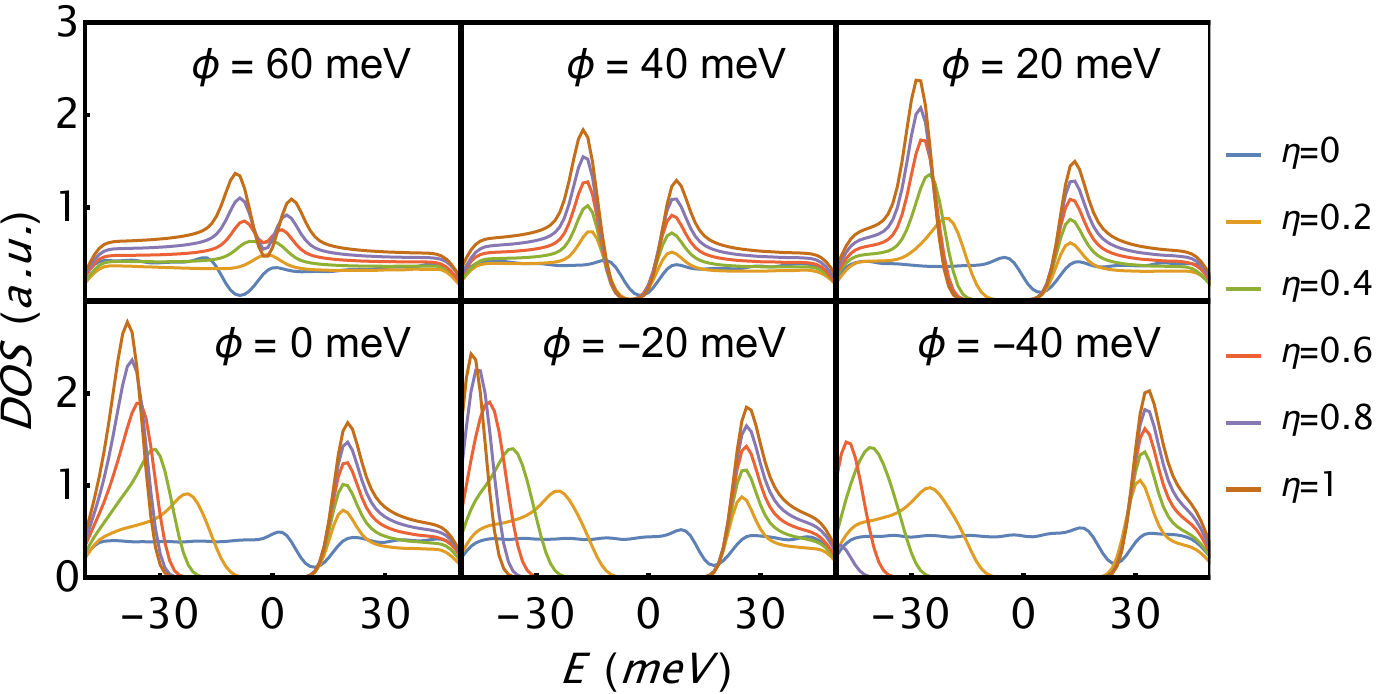}
    
    \caption{
        DOS of ABC islands using different area factors $\eta$ in the effective Hamiltonian.
        \textbf{(a)} DOS for different values of $\Delta_1$.
                     In general, decreasing $\eta$ suppresses the peaks.
                     An exception occurs for $\Delta_1 = 50 \, meV$ where until $\eta = 0.6$ the peaks actually increase as $\eta$ is decreased.
        \textbf{(b)} DOS for different values of $\phi$, with $\Delta_1 = 30 \, meV$.
                     We observe a close correspondence between the resulting DOS with the ones shown in (a), in accordance with~\Cref{eq:renormalized_onsite_potentials}.
    }
    \label{fig:abc_dos}
\end{figure}

We investigate the effects of the islands on the low-energy density of states (DOS) using the effective Hamiltonian of an ABC island for flavor-symmetric occupation.
\Cref{fig:abc_dos} shows the resulting DOS for different applied displacement fields $\Delta_1$ and relative island areas $\eta$.
In general, decreasing $\eta$ suppresses the DOS peaks.
This is to be expected because the effective band structure interpolates between the pristine ABC trilayer and the combined Bernal bilayer plus monolayer stack. 
Since the electron density and the trigonal warping generically increase with higher layer number~\cite{Koshino2009,Zhang2010}, an effective dispersion corresponding to an intermediate stacking number between two and three will have more dispersive bands.
For certain fine-tuned parameter combinations, the band flattening from the displacement field can reverse these tendencies. For example,
for $\Delta_1 = 50 \, meV$ we find effective band structures featuring more pronounced van Hove peaks than the pristine ABC trilayer.

It is known that between different graphene multilayers, there can be a large work function difference, reaching values up to $\sim 150 \, \mathrm{meV}$~\cite{Roy2025}. This can be taken into account in the model as additional relative layer potentials that differ from those of pristine ABC.
To be concrete, let us denote the additional layer potential of the top monolayer by $\phi$, which renormalizes the onsite potentials $\Delta_1$ and $\Delta_2$ according to \Cref{eq:abc_island:e},
\begin{equation}
    \Delta_1 \rightarrow \Delta_1 - \frac{\phi}{2}, \qquad
    \Delta_2 \rightarrow \Delta_2 + \frac{\phi}{6}.
    \label{eq:renormalized_onsite_potentials}
\end{equation}
$\phi<0$ can be interpreted as the addition of a constant layer potential to the bottom two layers.
\Cref{fig:abc_dos}b compares the low-energy DOS for different values of $\phi$.
For example, the DOS for $\phi = 40 \, \mathrm{meV}$ in \Cref{fig:abc_dos}b ($\Delta_1$ effectively becomes $10 \, \mathrm{meV}$) is comparable to the DOS for $\Delta_1 = 10 \, \mathrm{meV}$ in \Cref{fig:abc_dos}a.
Obviously, determining the actual difference in work function for a nanoscale island requires a self-consistent solution of the electrostatics, which we leave for future work. However, we note that since the different graphenes touch directly at the interface, the residual voltage differences are expected to be substantially below the bare values of isolated multilayers~\cite{Giovannetti2008}.

\paragraph{Conclusions.---}
\label{section:discussion}

We have investigated the electronic properties of graphene islands using a tight-binding description that explicitly accounts for the coupling between the island and its surrounding graphene contacts.
We find that the electron wavefunction is not confined to the island, but rather is distributed between the island and contacts in proportion to their relative areas, with the distribution remaining approximately uniform across the unit cells comprising the supercell.
As far as we could test, these band-structure modifications persist as the island size is increased, indicating an unusually large proximity length scale.
Consequently, the electronic structure of a nanoscale island cannot, in general, be understood simply as that of an isolated graphene stack. Instead, the island and contacts form a hybrid electronic system whose band structure interpolates between different layer stackings.
We proposed an effective single-unit-cell Hamiltonian that reproduces the low-energy bands of the supercell, where the effective hopping parameters scale with
the relative island area.

In our model calculations we did not modify the trigonal warping between island and leads, and the relative work function and onsite potentials at dissimilar graphene positions were not taken into account self-consistently. In the future, it will be worthwhile to add these ingredients in order to make predictions for many-layer stacking configurations yet to be realized experimentally.

Our results establish graphene islands as a novel type of band-engineered platform in which the electronic structure can be controlled by the island's nanoscale dimensions.
The area-averaging mechanism underlying the effective Hamiltonian may also be applicable to other vdW materials like TMDs or intercalated compounds~\cite{Manzeli2017,Nair2020,Zhai2026,Zhang2026}, opening the door to a continuous change of the layer number, even in the few-layer limit.

\begin{acknowledgments}
We thank 
Moshe Ben Shalom, 
Simon Salleh Atri,
Yoav Sharaby,
Diana Golovanova 
and Binghai Yan
for insightful discussions.
T.H.\ acknowledges financial support by the 
European Research Council (ERC) under grant QuantumCUSP
(Grant Agreement No. 101077020). 
\end{acknowledgments}

\paragraph{Data availability---} The data that support the findings of this article are available from the authors upon reasonable request.

%

\pagebreak
\onecolumngrid
\newpage
\appendix
\setcounter{page}{1}
\thispagestyle{empty}
\begin{center}
\textbf{\large Supplementary Information for\\ ``Graphene Islands as a Platform for Band-Structure Engineering"}
\\[6pt]
Yonah Mann, Tobias Holder

{\small 
School of Physics and Astronomy, Tel Aviv University, Tel Aviv 69978, Israel
}
\end{center}

In this Supplemental Information, we review the tight-binding Hamiltonians used in the main text for ABA, ABC and ABCA graphenes, give the details of the fitting for the effective island Hamiltonian, crosscheck the role of layer-dependent potentials and provide additional data regarding the spatial distribution of the charge density.

\renewcommand{\thefigure}{\arabic{figure}}
\renewcommand{\figurename}{Supplementary Figure}
\renewcommand{\tablename}{Supplementary Table}
\setcounter{figure}{0}
\renewcommand{\theHfigure}{S\arabic{figure}}
\renewcommand{\theHtable}{S\arabic{table}}

\Crefname{figure}{SFig{.}}{SFigs{.}}

\section{Pristine Tight-Binding Model}

\label{appendix:tight_binding_parameters}

For completeness, we summarize the pristine tight-binding Hamiltonians and parameter values used throughout this work, taken from \cite{Zibrov2018,Zhou2021,Zhou2023,Ghazaryan2023}. For the ABC and ABA trilayers, the Hamiltonians are,
\small{
\begin{subequations}\label{eq:pristine_trilayer}
\begin{align}
    H_{\text{ABC},0} &= 
    \begin{pmatrix}
    0 & v_0 \pi^\dagger & v_4 \pi^\dagger & v_3 \pi & 0 & \frac{1}{2}\gamma_2 \\
    v_0 \pi & \delta & \gamma_1 & v_4 \pi^\dagger & 0 & 0 \\
    v_4 \pi & \gamma_1 & \delta & v_0 \pi^\dagger & v_4 \pi^\dagger & v_3 \pi \\
    v_3 \pi^\dagger & v_4 \pi & v_0 \pi & \delta & \gamma_1 & v_4 \pi^\dagger \\
    0 & 0 & v_4 \pi & \gamma_1 & \delta & v_0 \pi^\dagger \\
    \frac{1}{2}\gamma_2 & 0 & v_3 \pi^\dagger & v_4 \pi & v_0 \pi & 0
    \end{pmatrix} \label{eq:pristine_trilayer:a} \\
    H_{\text{ABA},0} &= 
    \begin{pmatrix}
    0 & v_0 \pi^\dagger & v_4 \pi^\dagger & v_3 \pi & \frac{1}{2} \gamma_2& 0 \\
    v_0 \pi & \delta & \gamma_1 & v_4 \pi^\dagger & 0 & \frac{1}{2}\gamma_5 \\
    v_4 \pi & \gamma_1 & \delta & v_0 \pi^\dagger & v_4 \pi & \gamma_1 \\
    v_3 \pi^\dagger & v_4 \pi & v_0 \pi & 0 & v_3 \pi^\dagger & v_4 \pi \\
    \frac{1}{2}\gamma_2 & 0 & v_4 \pi^\dagger & v_3 \pi & 0 & v_0 \pi^\dagger \\
    0 & \frac{1}{2}\gamma_5 & \gamma_1 & v_4 \pi^\dagger & v_0 \pi & \delta
    \end{pmatrix} \label{eq:pristine_trilayer:b} \\
    H_{\text{ABC}} &= H_{\text{ABC},0} + H_{\text{tri},\Delta_1} + H_{\text{tri},\Delta_2} \label{eq:pristine_trilayer:c} \\
    H_{\text{ABA}} &= H_{\text{ABA},0} + H_{\text{tri},\Delta_1} + H_{\text{tri},\Delta_2} \label{eq:pristine_trilayer:d}
\end{align}
\end{subequations}
}
where $H_{\text{tri},\Delta_1}$ and $H_{\text{tri},\Delta_2}$ are defined 
as
\begin{align}
    H_{\text{tri},\Delta_1} &= \operatorname{diag}(\Delta_1, \Delta_1, 0, 0, -\Delta_1, -\Delta_1) \\
    H_{\text{tri},\Delta_2} &= \operatorname{diag}(\Delta_2, \Delta_2,
    -2\Delta_2, -2\Delta_2,
     \Delta_2, \Delta_2).
\end{align}
The ABCA tetralayer can be written as,
\small{
\begin{subequations}\label{eq:pristine_abca}
\begin{align}
    H_{\text{ABCA},\Delta_1} &= diag(\Delta_1, \Delta_1, \frac{\Delta_1}{3},
                              \frac{\Delta_1}{3}, -\frac{\Delta_1}{3},
                              -\frac{\Delta_1}{3},-\Delta_1, -\Delta_1) \label{eq:pristine_abca:a} \\
    H_{\text{ABCA},\Delta_2} &= diag(\Delta_2, \Delta_2,
                              -\Delta_2, -\Delta_2,
                              -\Delta_2, -\Delta_2,
                              \Delta_2, \Delta_2) \label{eq:pristine_abca:b} \\
    H_{\text{ABCA},0} &=
    \begin{pmatrix}
        0 & v_0 \pi^\dagger & v_4 \pi^\dagger & v_3 \pi & 0 & \frac{1}{2}\gamma_2 & 0 & 0 \\
        v_0 \pi & \delta & \gamma_1 & v_4 \pi^\dagger & 0 & 0 & 0 & 0 \\
        v_4 \pi & \gamma_1 & \delta & v_0 \pi^\dagger & v_4 \pi^\dagger & v_3 \pi & 0 & \frac{1}{2}\gamma_2 \\
        v_3 \pi^\dagger & v_4 \pi & v_0 \pi & \delta & \gamma_1 & v_4 \pi^\dagger & 0 & 0 \\
        0 & 0 & v_4 \pi & \gamma_1 & \delta & v_0 \pi^\dagger & v_4 \pi^\dagger & v_3 \pi \\
        \frac{1}{2}\gamma_2 & 0 & v_3 \pi^\dagger & v_4 \pi & v_0 \pi & \delta & \gamma_1 & v_4 \pi^\dagger \\
        0 & 0 & 0 & 0 & v_4 \pi & \gamma_1 & \delta & v_0 \pi^\dagger \\
        0 & 0 & \frac{1}{2}\gamma_2 & 0 & v_3 \pi^\dagger & v_4 \pi & v_0 \pi & 0
    \end{pmatrix} \label{eq:pristine_abca:c} \\
    H_{\text{ABCA}} &= H_{\text{ABCA},0} + H_{\text{ABCA},\Delta_1} + H_{\text{ABCA},\Delta_2} .\label{eq:pristine_abca:d}
\end{align}
\end{subequations}
}
The pristine tight-binding models have the hopping parameters shown in Table I and were validated by comparing with the energy bands and Fermi surfaces in \cite{Zibrov2018,Zhou2021,Zhou2023,Ghazaryan2023}.

\begin{figure}[h]
\begin{center}
    \makebox[.5\columnwidth][l]{%
        \panelimage[box=0.16\columnwidth,
                    title=ABA,
                    titlexshift=0.5em,
                    titleyshift=0.5em,
                    graphicsopts={width=0.16\columnwidth, trim=0.1cm 0 0.1cm 0, clip}]{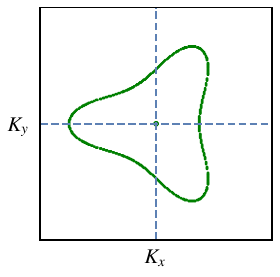}
        \panelimage[box=0.16\columnwidth,
                    title=ABC,
                    titlexshift=0.5em,
                    titleyshift=0.5em,
                    graphicsopts={width=0.16\columnwidth, trim=0.1cm 0 0.1cm 0, clip}]{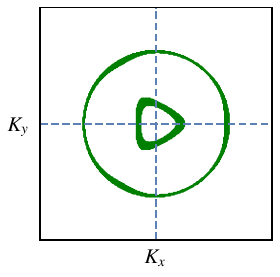}
        \panelimage[box=0.16\columnwidth,
                    title=ABCA,
                    titlexshift=0.5em,
                    titleyshift=0.5em,
                    graphicsopts={width=0.16\columnwidth, trim=0.1cm 0 0.1cm 0, clip}]{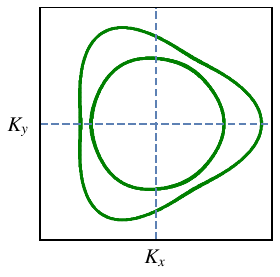}
    }
    \makebox[.5\columnwidth][l]{%
        \panelimage[box=0.16\columnwidth,
                    graphicsopts={width=0.16\columnwidth, trim=0.1cm 0 0.1cm 0, clip}]{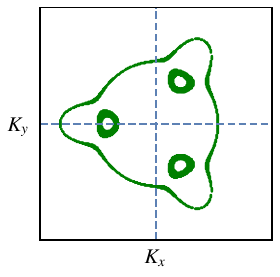}
        \panelimage[box=0.16\columnwidth,
                    graphicsopts={width=0.16\columnwidth, trim=0.1cm 0 0.1cm 0, clip}]{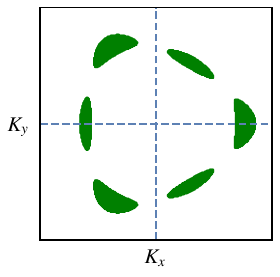}
        \panelimage[box=0.16\columnwidth,
                    graphicsopts={width=0.16\columnwidth, trim=0.1cm 0 0.1cm 0, clip}]{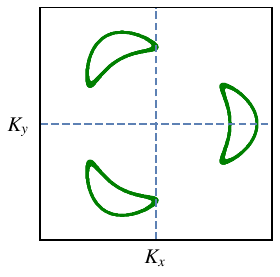}
    }
\end{center}
    \caption{
        Fermi surfaces of pristine tight-binding Hamiltonians for different stackings.
        The agreement with known band structures and Fermi surfaces \cite{Zibrov2018,Zhou2021,Zhou2023,Ghazaryan2023} validates the implementation of the pristine tight-binding models used throughout this work.
    }
    \label{fig:pristine_fermi_surfaces}
\end{figure}

\begin{table}[!ht]
\setlength{\tabcolsep}{4pt}
\centering
\begin{tabular}{lccccccccc}
\toprule
 &
$\gamma_0$ &
$\gamma_1$ &
$\gamma_2$ &
$\gamma_3$ &
$\gamma_4$ &
$\gamma_5$ &
$\delta$ &
$\Delta_2$ \\
\midrule
ABA  & 3100 & 380 & -21 & \phantom{-}290  & \phantom{-}141  & 50 & \phantom{-}35.5  & \phantom{-}3.5  \\
ABC  & 3100 & 380 & -15 & -290 & -141 &    & -10.5 & -2.3 \\
ABCA & 3100 & 380 & -15 & -290 & -141 &    & -10.5 & 0    \\
\bottomrule
\end{tabular}
\caption{
    Tight-binding parameters (in meV) used for each stacking configuration.
    The magnitude of the hopping amplitudes are nearly independent of the stacking configuration.
    Sign differences reflect the chosen convention.
    }
\label{tab:tight_binding_parameters}
\end{table}

\section{Effective Island Hamiltonian}
\label{appendix:effective_island_hamiltonian}

\begin{figure}
    \includegraphics[width=0.65\columnwidth, clip]{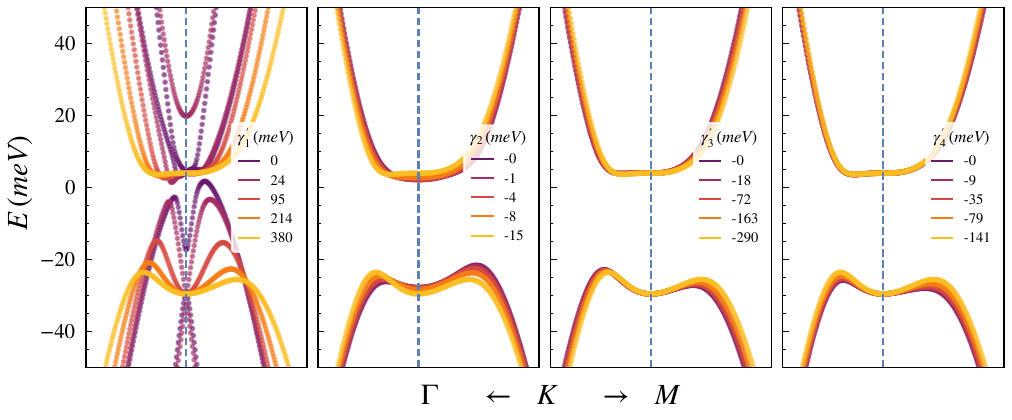}
    \caption{
        Sweep of modified parameter $\gamma_1'$, $\gamma_2$, $\gamma_3'$, and $\gamma_4'$ for an ABC island ($\Delta_1 = 15 \, meV$), while keeping the other parameters the same value as the pristine stacking of the island.
        Although $\gamma_2$, $\gamma_3'$, and $\gamma_4'$ do have an effect on the band structure, $\gamma_1'$ is the modified parameter which controls how much the band structure is similar to ABC graphene vs how much it is similar to MLG and BLG.
    }
    \label{fig:abc_modified_parameter_sweep}
\end{figure}

In the main text, we derived an effective pristine Hamiltonian for an ABC island.
We first examine the sensitivity of the effective band structure to the modified hopping parameters.
We then use this information to develop a procedure for extracting these parameters from the full supercell calculation.
\Cref{fig:abc_modified_parameter_sweep} depicts the evolution of the band structures when sweeping the value of a modified hopping amplitude, while keeping the rest the same as the pristine value.
This shows that the evolution of the TLG low-energy band structure toward the MLG and BLG limit is primarily controlled by $\gamma_1'$.
The remaining modified parameters produce smaller changes in the low-energy bands.
Another important factor to notice is that only $\gamma_1'$ and $\gamma_2'$ affect the low-energy bands at the $K$ point.
This is expected, since $\gamma_3'$ and $\gamma_4'$ terms correspond to momentum-dependent hopping that vanish at the $K$ point.

This motivates us to attempt to extract $\gamma_1'$ and $\gamma_2'$ using the energy-band values at the $K$ point.
The three pairs of energies at the $K$ point are calculated by setting $\pi = 0$ in \Cref{eq:abc_island}:
\begin{subequations}
\begin{align}
    \delta + \Delta_2 &\pm \frac{1}{2} \sqrt{\gamma _2{}^2+4 \Delta _1^2} \\
    \frac{\Delta_1 - \Delta_2}{2} &\pm \frac{1}{2}\sqrt{4 \gamma _1{}^2+\left(\Delta _1 + 3 \Delta _2\right){}^2} \\
    -\frac{\Delta_1 + \Delta_2}{2} &\pm \frac{1}{2}\sqrt{4 \gamma _1'{}^2+\left(\Delta _1 - 3 \Delta _2\right){}^2}.
\end{align}
    \label{eq:k_point_energies_abc}
\end{subequations}
As expected, $\gamma_3'$ and $\gamma_4'$ do not appear in the $K$-point energies.
By numerically calculating the $K$-point energies of the supercell, we utilize \Cref{eq:k_point_energies_abc} to extract the values of $\gamma_1'$ and $\gamma_2'$.
The remaining parameters, $\gamma_3'$ and $\gamma_4'$, are determined by minimizing the deviation between the effective and supercell low-energy spectra.
Specifically, we minimize the loss function
\begin{equation}
    L = \sqrt{\frac{1}{m} \sum_{i=1}^m \left(E_i^{\text{eff}} - E_i^{\text{sc}}\right)^2} + \alpha \left|N_{\text{eff}} - N_{\text{sc}}\right|
\end{equation}
where $E_i^{\text{eff}}$ and $E_i^{\text{sc}}$ denote the $i$-th ordered low-energy eigenvalues of the effective and full-supercell Hamiltonians, respectively.
Here $N_{\text{eff}}$ and $N_{\text{sc}}$ are the number of low-energy bands in the effective model and the supercell model respectively, $m = \text{min}\left(N_{\text{eff}}, N_{\text{sc}}\right)$, and $\alpha$ is a dimensionless parameter.
The first term is the root-mean-square deviation between the ordered eigenvalues \cite{Hodson2022}, while the second term penalizes a mismatch in the number of low-energy eigenvalues included in the comparison.
We chose $\alpha = 10$, which resulted in stable fits.
Since \Cref{fig:abc_modified_parameter_sweep} shows that variations in $\gamma_3'$ and $\gamma_4'$ have only a minor effect on the low-energy band structure, these parameters are weakly constrained by the fitting procedure.
It is therefore expected that their fitted values do not perfectly follow the relation $\gamma'=\eta\gamma$ [Eq. (5) of the main text].

We now show that the results of this procedure are also valid for other stacking configurations.
The effective Hamiltonians of ABA and ABCA islands are constructed by modifying the pristine Hamiltonians defined in \Cref{eq:pristine_trilayer,eq:pristine_abca} in the same manner as was done for ABC in \Cref{eq:abc_island}:
\small{
\begin{subequations}\label{eq:aba_island}
\begin{align}
    H_{\text{ABA},0}^{\text{island}} &= 
    \begin{pmatrix}
    0 & v_0 \pi^\dagger & v_4 \pi^\dagger & v_3 \pi & \frac{1}{2} \gamma_2'& 0 \\
    v_0 \pi & \delta & \gamma_1 & v_4 \pi^\dagger & 0 & \frac{1}{2}\gamma_5' \\
    v_4 \pi & \gamma_1 & \delta & v_0 \pi^\dagger & v_4' \pi & \gamma_1' \\
    v_3 \pi^\dagger & v_4 \pi & v_0 \pi & 0 & v_3' \pi^\dagger & v_4' \pi \\
    \frac{1}{2}\gamma_2' & 0 & v_4' \pi^\dagger & v_3' \pi & 0 & v_0 \pi^\dagger \\
    0 & \frac{1}{2}\gamma_5' & \gamma_1' & v_4' \pi^\dagger & v_0 \pi & \delta
    \end{pmatrix} \label{eq:aba_island:a} \\
    H_{\text{ABA}}^{\text{island}} &= H_{\text{ABA},0}^{\text{island}} + H_{\text{tri},\Delta_1} + H_{\text{tri},\Delta_2} \label{eq:aba_island:b}
\end{align}
\end{subequations}
}
where $H_{\text{tri},\Delta_1}$ and $H_{\text{tri},\Delta_2}$ are defined in \Cref{eq:abc_island}.

\begin{figure}[h]
\includegraphics[width=.55\columnwidth]{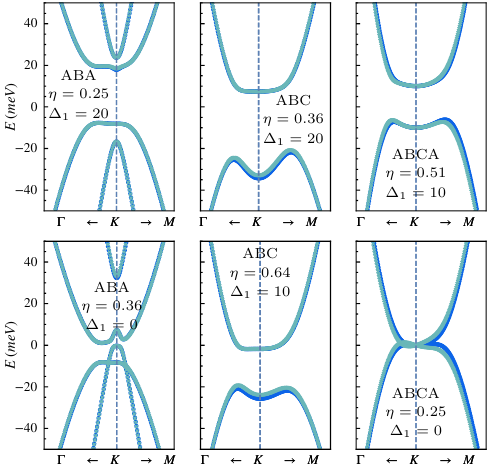}

    \caption{
        Comparison of energy bands calculated using the full supercell (dark blue) vs using the effective Hamiltonian (light blue) for different stacking configurations, relative island areas $\eta$, and applied displacement fields $\Delta_1$ (in meV).
    }
    \label{fig:supercell_vs_effective_hamiltonian_energy_bands}
\end{figure}

\small{
\begin{subequations}\label{eq:abca_island}
\begin{align}
    H_{\text{ABCA},0}^{\text{island}} &=
    \begin{pmatrix}
        0 & v_0 \pi^\dagger & v_4 \pi^\dagger & v_3 \pi & 0 & \frac{\gamma_2'}{2} & 0 & 0 \\
        v_0 \pi & \delta & \gamma_1 & v_4 \pi^\dagger & 0 & 0 & 0 & 0 \\
        v_4 \pi & \gamma_1 & \delta & v_0 \pi^\dagger & v_4' \pi^\dagger & v_3' \pi & 0 & \frac{\gamma_2'}{2} \\
        v_3 \pi^\dagger & v_4 \pi & v_0 \pi & \delta & \gamma_1' & v_4' \pi^\dagger & 0 & 0 \\
        0 & 0 & v_4' \pi & \gamma_1' & \delta & v_0 \pi^\dagger & v_4 \pi^\dagger & v_3 \pi \\
        \frac{\gamma_2'}{2} & 0 & v_3' \pi^\dagger & v_4' \pi & v_0 \pi & \delta & \gamma_1 & v_4 \pi^\dagger \\
        0 & 0 & 0 & 0 & v_4 \pi & \gamma_1 & \delta & v_0 \pi^\dagger \\
        0 & 0 & \frac{\gamma_2'}{2} & 0 & v_3 \pi^\dagger & v_4 \pi & v_0 \pi & 0
    \end{pmatrix} \label{eq:abca_island:a} \\
    H_{\text{ABCA}}^{\text{island}} &= H_{\text{ABCA},0}^{\text{island}} + H_{\text{ABCA},\Delta_1} + H_{\text{ABCA},\Delta_2} \label{eq:abca_island:b}
\end{align}
\end{subequations}
}
where $H_{\text{ABCA},\Delta_1}$ and $H_{\text{ABCA},\Delta_2}$ are defined in \Cref{eq:pristine_abca}.
Notice that for ABA islands there is an additional modified parameter $\gamma_5'$.
We apply the same fitting procedure described above, using \Cref{eq:modified_vs_pristine_hopping_ratio} to initialize the modified hopping parameters.
\Cref{fig:supercell_vs_effective_hamiltonian_energy_bands} compares the resulting energy bands with the ones calculated using the full supercell, and demonstrates that the effective Hamiltonian accurately reproduces the low-energy bands of the full supercell.

\section{Validation of Contact Simplification}
\label{appendix:consequence_of_contact_onsite_potential_incompatibility}

\begin{figure}[h]
    \includegraphics{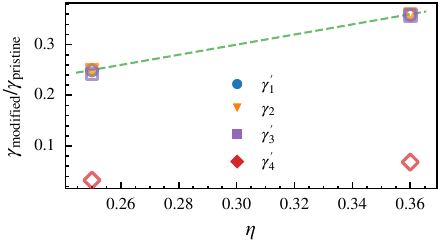}
    \caption{
        Same as Fig.~4 of the main text, but with the corrected contact onsite potentials, for $\Delta_1 = 10 \, meV$.
        The fitted amplitudes continue to follow Eq.~(5) of the main text.
    }
    \label{fig:modified_parameters_with_modified_onsite_split}
\end{figure}

In the main text we focused on a simplification in the construction of the island model, where
interlayer hoppings across the spacer were removed, while the remaining hopping amplitudes as well as the onsite potentials were kept as those of the pristine island stacking.
As a result, the onsite potentials in the contacts do not correspond to those of the actual contact stacking.

To determine whether this approximation affects the effective Hamiltonian, we repeat the fitting procedure described above using the corrected contact onsite potentials.
The resulting modified hopping amplitudes are shown in \Cref{fig:modified_parameters_with_modified_onsite_split}.
The fitted hopping amplitudes continue to satisfy $\gamma'=\eta\gamma$.
This demonstrates that the assumption of uniform onsite potentials  does not qualitatively affect the  effective Hamiltonian.

The only exception was the onsite potential of the dimer orbital corresponding to the monolayer component of the contacts in the ABA and ABC effective Hamiltonians.
In order to correctly match the bands of the supercell calculation with those of the effective Hamiltonian, an additional onsite potential correction to the isolated dimer orbital was required.
When the contact onsite potentials were inserted in the first place, this correction became unnecessary.
We therefore attribute this correction to the artificial onsite potential mismatch introduced by the simplified contact model, rather than to any physical property of the graphene islands.

Another result of using this approximation is that the relative work functions of the different stackings do not appear in our model \cite{Roy2025}.
When applying a displacement field, we are in effect introducing a work function between the island and the layers in the contact.
The observed independence of our results on the applied displacement field therefore very strongly indicates that the addition of relative work function differences is unlikely to qualitatively affect our conclusions.

\section{Wavefunction Distributions}
\label{appendix:wavefunction_distribution}

\begin{figure}
    \makebox[.5\columnwidth][l]{%
        \hspace{-1.18em}
        \panelimage[box=0.25\columnwidth,
                    graphicsopts={height=5cm, trim=0 0 1.7cm 0, clip}]{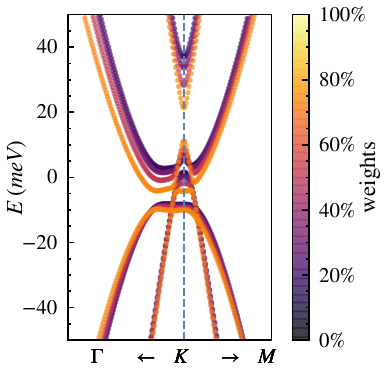}
        \hspace{-2em}
                \panelimage[box=0.25\columnwidth,
                            graphicsopts={height=5cm, trim=1cm 0 0.1 0, clip}]{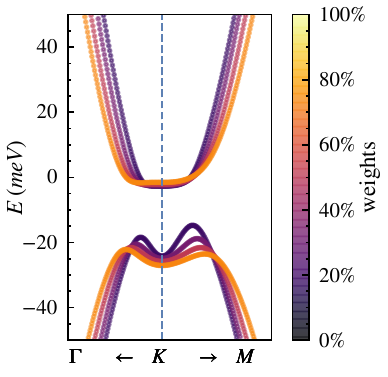}
    }%
 
    \caption{
        Same as Fig.~3 of the main text, but for representative ABA $\Delta_1 = 0$ (left) and ABC $\Delta_1 = 10 \, meV$(right) islands.
        This demonstrates that the hybridization is independent of stacking configuration and applied displacement field.
        }
        
    \label{fig:wavefunction_k_space_distribution_other_islands}
\end{figure}

\begin{figure}
    \includegraphics{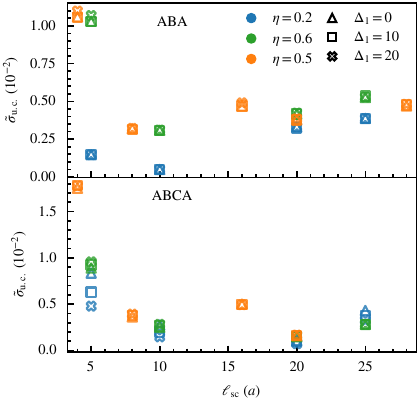}
    \caption{
        Same as Fig.~3 for ABA (top) and ABCA (bottom) islands.
        The weight of the wavefunction across the different unit cells remains uniform regardless of stacking configuration, even as supercell size grows.
        }
    \label{fig:wavefunction_unit_cell_deviation_others}
\end{figure}

In the main text we showed that for an ABCA island, the wavefunction is evenly spread out, independently of momentum, applied displacement field, and relative area of the island.
Furthermore, we showed that for an ABC island the wavefunction is evenly distributed between the different unit cells of the supercell.
Here we demonstrate that this behavior is not specific to a particular stacking configuration.

\Cref{fig:wavefunction_k_space_distribution_other_islands} shows the wavefunction weight for representative ABA and ABC islands.
As in Fig.~3 of the main text, the weight is independent of momentum near the $K$ point.
\Cref{fig:wavefunction_unit_cell_deviation_others} shows the normalized standard deviation as defined in the main text for representative ABA and ABCA islands.
As in Fig.~4 of the main text, the wavefunction is uniformly distributed among the unit cells in the supercell, even when the supercell is enlarged.

\begin{figure}
    \includegraphics{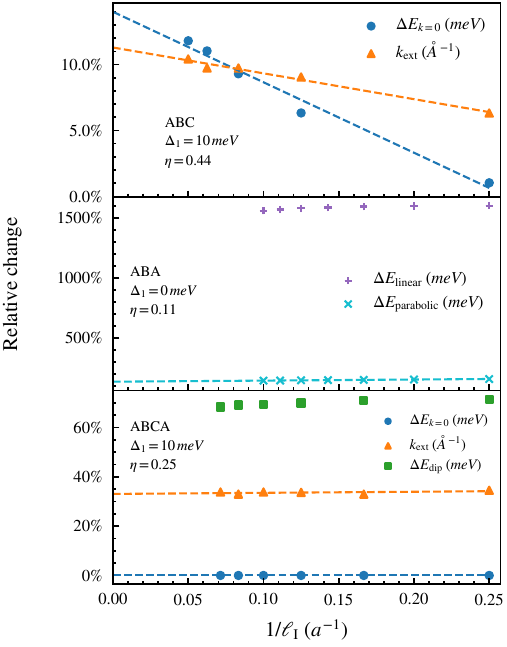}
    \caption{
        Scaling of band-structure features as a function of system size.
        The relative area ($\eta$) is kept constant, while the island size ($\ell_\text{I}$) is increased.
        Linear fits are shown for features exhibiting a linear dependence on $1/\ell_\text{I}$.
        Notably, many of these modifications remain finite, demonstrating that they are not merely finite-size effects.
        }
    \label{fig:modified_features}
\end{figure}

We next investigate whether the band-structure modifications persist as the island size is increased, rather than arising solely from finite-size effects.
To do so we look at the scaling of the modifications of several features, where the modification is quantified as the relative deviation from the pristine value:
\begin{equation}
    \left|\frac{y_s - y_{\text{pristine}}}{y_{\text{pristine}}}\right|
\end{equation}
where $y_s$ denotes a band-structure feature of the island and $y_{\text{pristine}}$ the corresponding pristine value.

For ABC islands we explore the energy gap at the $K$ point $\Delta E_{k=0}$ and the trigonal warping $k_{\text{ext}}$, where $k_{\text{ext}}$ is defined as the momentum of the leftmost extremum.
For ABA islands we explore the energy gaps of the linear $\Delta E_{\text{linear}}$ and parabolic $\Delta E_{\text{parabolic}}$ bands.
For ABCA island we explore the energy gap at the $K$ point $\Delta E_{k=0}$, the trigonal warping $k_{\text{ext}}$, the energy gap at the leftmost extremum $\Delta E_{\text{dip}}$.
    
\Cref{fig:modified_features} depicts the scaling of these features as a function of inverse island size $\ell_\text{I}$.
For several of the explored features, a non-vanishing intercept is predicted at infinite size, indicating that the band-structure modifications do not vanish exponentially in system size, as one would expect for a regular proximity effect.
It is important to note that the present description is valid only over length scales smaller than the quantum coherence length, beyond which the spatially separated island and contact regions may be treated as effectively separable~\cite{Malpetti2016}.
Beyond such length scales, we therefore expect the band-structure features to approach those of the corresponding pristine graphene system, irrespective of the present calculation, which implicitly assumes an infinite quasiparticle lifetime.

\end{document}